\documentclass[aps,prl,twocolumn,groupedaddress,superscriptaddress]{revtex4}

\usepackage{graphicx}
\usepackage{hyperref}
\usepackage{dcolumn}
\usepackage{bm}
\usepackage{amssymb,amsmath}

\newcommand{\vs}{\vspace*}
\newcommand{\np}{\newpage}
\newcommand{\w}{\omega}

\newcommand{\Eref}[1]{Eq.~(\ref{#1})}
\newcommand{\Fref}[1]{Fig.~\ref{#1}}

\newcommand{\be}{\begin{equation}}
\newcommand{\ee}{\end{equation}}
\newcommand{\br}{\begin{eqnarray*}}
\newcommand{\er}{\end{eqnarray*}}
\newcommand{\ba}{\begin{eqnarray}}
\newcommand{\ea}{\end{eqnarray}}
\newcommand{\bp}{\begin{minipage}}
\newcommand{\ep}{\end{minipage}}
\newcommand{\bt}{\begin{tabular}}
\newcommand{\et}{\end{tabular}}

\newcommand{\ms}{\vspace*{-5mm}}
\newcommand{\mms}{\vspace*{-2.5mm}}

\graphicspath{{./}{./PDF/}{./EPS/}{./PNG/}{./Data/}}

\begin{document}
\bibliographystyle{apsrev}


\title{Rainbow RABBITT as a Probe of Coherent Rabi Dynamics}

\author{Vladislav~V.~Serov}
\affiliation{Department of Medical Physics, Saratov State University,
             Saratov 410012, Russia}

\author{Anatoli~S.~Kheifets}
\affiliation{Research School of Physics, Australian National University,
             Canberra ACT 2601, Australia}

\begin{abstract}
Attosecond pulse trains interacting with a resonantly dressed atom
generate a pronounced intra-sideband phase structure that remains
hidden in conventional spectrally integrated RABBITT
measurements. Using \textit{ab initio} time-dependent Schr\"odinger
equation calculations for lithium near the resonant $2s\to2p$
transition, we show that the phase extracted within a single sideband
can vary by nearly $\pi$ across its spectral width. The resulting
intra-sideband phase dispersion exhibits a characteristic dependence
on the IR detuning, pulse duration, intensity, and sideband order. Most
strikingly, exact resonant Rabi flopping flattens the intra-sideband
phase dispersion, whereas a small detuning generates a pronounced
phase modulation despite weaker population transfer. This
counterintuitive behavior demonstrates that rainbow RABBITT probes the
dynamical phase accumulated by a Rabi-dressed wave packet rather than
the instantaneous populations of the participating states.  A simple
analytical model captures the principal features of the numerical
calculations and provides physical insight into the emergence of the
intra-sideband phase structure.  These results establish
intra-sideband phase dispersion as a new interferometric observable
for mapping coherent Rabi dynamics.
\end{abstract}

\maketitle
\ms
\paragraph*{Introduction.}

Coherent control and direct observation of ultrafast quantum dynamics
are central goals of attosecond science, recognized by the 2023 Nobel
Prize in Physics~\cite{LHuillier2024,Agostini2024,Krausz2024}.  Recent
advances in attosecond interferometry and phase retrieval have enabled
access to the spectral phase and coherent structure of electronic wave
packets in atoms, molecules, and
solids~\cite{Pedrelli2020,Berkane2025,Fuchs2021}.  At the same time,
resonantly driven quantum systems exhibit coherent dressed-state
dynamics, including Rabi oscillations, population transfer, and
dynamical phase accumulation~\cite{AllenEberly1975}. Such dynamics
have been observed at XUV and free-electron-laser wavelengths in
helium~\cite{Nandi2022,Nandi2024} and at near-infrared wavelengths in
lithium~\cite{Liao2022}. Because interferometric observables can
differ qualitatively from population-based observables
\cite{Gaynor2023,Jakob2025,Rico2025}, phase-sensitive probes provide
access to coherent light--matter dynamics that are not directly
encoded in population transfer alone.

Reconstruction of attosecond beating by interference of two-photon
transitions (RABBITT)~\cite{Muller2002,TomaJPB2002} is a cornerstone
of attosecond chronoscopy~\cite{RevModPhys.87.765}. In conventional
RABBITT, an attosecond pulse train (APT) ionizes the target while a
dressing IR photon is either absorbed or emitted, allowing two quantum
pathways to reach the same continuum state (see Fig.~1a).  The
resulting sideband (SB) signal oscillates with the XUV--IR delay
$\tau$ as $S_{2q}(\tau)=A+B\cos(2\omega\tau+C)$, where $\omega$ is the
IR carrier frequency. The extracted phase
$C=\Delta\phi_{2q\pm1}+\Delta\phi_W+\Delta\phi_{cc}$ contains
contributions from the harmonic group-delay difference, the Wigner
photoionization phase, and the continuum--continuum (CC)
phase~\cite{Dahlstrom2012}. Its energy derivative yields the atomic
photoionization time delay.

\begin{figure}[t]
\ms
\includegraphics[width=0.8\columnwidth]{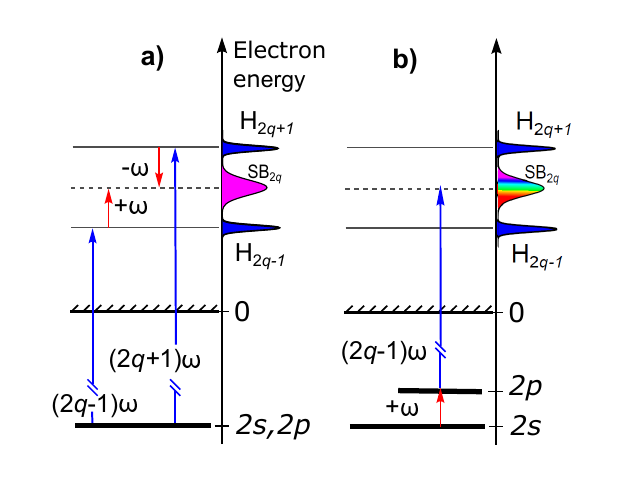}
\ms
\caption{
Schematic illustration of conventional (a) and
resonant (b)  RABBITT processes on Li. The rainbow phase retrieval is
highlighted in the latter case. 
 }
\label{Fig1}
\ms
\end{figure}

This interpretation changes qualitatively when the IR frequency is
tuned close to a bound-state resonance such as the
$2s\!\rightarrow\!2p$ transition in lithium. As illustrated in
\Fref{Fig1}b, the resonant IR field populates the intermediate $2p$
state and opens an additional ionization pathway that carries a
resonant phase instead of the usual CC
phase~\cite{PhysRevA.103.L011101,PhysRevA.104.L021103}. Because this
contribution is not associated with continuum--continuum
scattering, the conventional interpretation of $C(E)$ as a measure of
photoionization delay is no longer valid. Related strong-field studies
have revealed Autler--Townes splitting of sidebands and angle-resolved
phase structures arising from Rabi cycling of the initial
state~\cite{Liao2022,Mao2023,Liao2024}.

In the present work we extend the concept of spectrally
resolved, or rainbow, RABBITT analysis, previously used
to investigate spectral phase variations associated with
resonances, autoionization, and angle-dependent
photoionization
dynamics~\cite{Fuchs2021,Kotur2016,Gruson734,Busto2018,Isinger2019,Turconi2020,Neoricic2022,Roantree2023}.
Rather than integrating the sideband yield, we
extract the phase $C(E)$ independently at each
photoelectron energy within the sideband. This
rainbow decomposition reveals a pronounced
intra-sideband (ISB) phase structure hidden by the
conventional sideband-integrated, or extra-sideband
(ESB), phase.

Applying this approach to lithium near the resonant
$2s\to2p$ transition, we uncover a direct connection
between the ISB phase and the coherent dynamics of
a Rabi-dressed bound-state wave packet. The numerical
results indicate that the ISB phase is primarily sensitive
to the dynamical phase accumulated during the Rabi
cycle rather than to the instantaneous populations of
the participating states.

As a result, its modulation reflects the coherent phase evolution of
the dressed state rather than population transfer.  It is strongest at
finite detuning, decreases for long IR pulses, and diminishes at exact
resonance despite maximal population transfer.  We support this
interpretation with an analytical model and \textit{ab initio} TDSE
calculations. The resulting ISB phase emerges as a new interferometric
observable for mapping coherent Rabi dynamics through spectrally
resolved photoelectron interferometry.

\paragraph*{Theoretical framework.}

The resonantly coupled $2s$--$2p$ system is described within the
rotating-wave approximation (RWA)~\cite{AllenEberly1975} by the
two-level Hamiltonian
\begin{equation}
H_{2s,2p}
=
\frac{1}{2}
\begin{pmatrix}
-\Delta & \Omega(t)\\
 \Omega(t) & \Delta
\end{pmatrix},
\label{eq:twolvl}
\end{equation}
where $\Delta=\omega-\omega_{sp}$ is the IR detuning from the
$2s\to2p$ transition, $\Omega(t)=d_{sp}\mathcal{E}_{\rm IR}(t)$
is the instantaneous Rabi frequency, and
$\Omega_D=\sqrt{\Omega^2+\Delta^2}$ is the dressed Rabi frequency.
For a constant-envelope IR field and an atom initially in the $2s$
state, the RWA yields
\begin{align}
a_s(t)&=\cos\frac{\Omega_D t}{2}
        +i\frac{\Delta}{\Omega_D}\sin\frac{\Omega_D t}{2},
\label{eq:as}\\
a_p(t)&=-i\frac{\Omega}{\Omega_D}\sin\frac{\Omega_D t}{2}.
\label{eq:ap}
\end{align}

The dynamical phase of the ground-state amplitude is
\begin{equation}
\varphi_s(t)=\arg a_s(t)
 =\arctan\!\left[\frac{\Delta}{\Omega_D}
 \tan\frac{\Omega_D t}{2}\right].
\label{eq:argAs}
\end{equation}
At exact resonance ($\Delta=0$), $\varphi_s(t)\equiv0$ despite
complete population transfer. Finite detuning generates oscillations
of $\varphi_s(t)$ whose amplitude is maximized near
$|\Delta|\sim\Omega$.

The two-photon ionization amplitude for sideband $2q$ at
photoelectron energy $E$ and XUV--IR delay $\tau$ can be written as
\begin{equation}
M(E,\tau)=M_{nr}^{(+)}(E)e^{+i\omega\tau}
         +M_{nr}^{(-)}(E)e^{-i\omega\tau}
         +M_r(E,\tau;\Delta).
\label{eq:Mtot}
\end{equation}
The first two terms represent the conventional RABBITT pathways.
The third term, $M_r$, describes ionization from the transiently
populated $2p$ state,
\begin{equation}
M_r(E,\tau;\Delta)\propto
\langle E|\hat d|2p\rangle
\int_{-\infty}^{\infty}
dt\,f_{\rm APT}(t-\tau)\,
a_p(t)\,
e^{i(E-\omega_x)t}.
\label{eq:Mr}
\end{equation}
where $\omega_x$ is the central XUV frequency.  \Eref{eq:Mr} shows
that the resonant contribution is the coherent Fourier sum of the
ionization amplitudes generated by successive APT pulselets, weighted
by the time-dependent excited-state amplitude $a_p(t)$.


At exact resonance,
$a_p(t)=-i\sin(\Omega t/2)$ and
$a_s(t)=\cos(\Omega t/2)$ maintain a fixed
relative phase of $-\pi/2$. In this limit, the
resonant dressing acts approximately as a common
time-dependent gate on the two RABBITT pathways,
so that
$
M^{(+)}_{\rm eff} M^{(-)*}_{\rm eff}
\propto |f(\tau)|^2
M^{(+)}_{\rm nr} M^{(-)*}_{\rm nr}.
$

For finite detuning, the dynamical phase $\varphi_s(t)$ breaks the
common-mode symmetry discussed above. Successive APT pulselets probe
the atom at different stages of the Rabi cycle, so that the resonant
contribution becomes the coherent sum
\begin{equation}
M_r^{f}(E;\Delta,T_{\rm IR})
\propto
\int_{-\infty}^{\infty}
dt\,f_{\rm APT}(t)\,
a_p(t)\,
e^{i(E-E_{2q})t}.
\label{eq:Mrf}
\end{equation}
The intra-sideband phase is then
\begin{equation}
\delta C(E)
=
\arg\!\left[M_r^{f}(E;\Delta,T_{\rm IR})\right]
-
\arg\!\left[M_r^{f(0)}(E;T_{\rm IR})\right],
\label{eq:dC}
\end{equation}
where $M_r^{f(0)}$ is evaluated with $a_p\to1$, corresponding to a
non-resonant reference.

\Eref{eq:dC} reduces to the static resonant phase
$\phi_r=\arctan(\Gamma_{\rm IR}/\Delta)$ of
Ref.~\cite{PhysRevA.104.L021103} in the monochromatic
limit $T_{\rm IR}\to\infty$. For finite pulses, however,
the Fourier transform retains information about the full
Rabi evolution. When $T_{\rm IR}\gg\tau_{\rm APT}$, all
pulselets sample nearly the same value of $a_p$,
producing a flat ISB phase profile. In contrast, for
$T_{\rm IR}\sim\tau_{\rm APT}$ the pulselets probe
different stages of the Rabi cycle, generating a
dispersive phase profile. The crossover between these
limits is governed by $\Omega_D$ and $T_{\rm IR}$.

\paragraph*{Computational method.}

The numerical calculations were performed by solving the
time-dependent Schr\"odinger equation for lithium in the
single-active-electron approximation. Details of the
atomic model, laser parameters, numerical propagation,
and rainbow-phase retrieval procedure are provided in the
Supplemental Material.

\begin{figure}[t]
\hspace*{6mm}

\includegraphics[width=1.0\columnwidth]{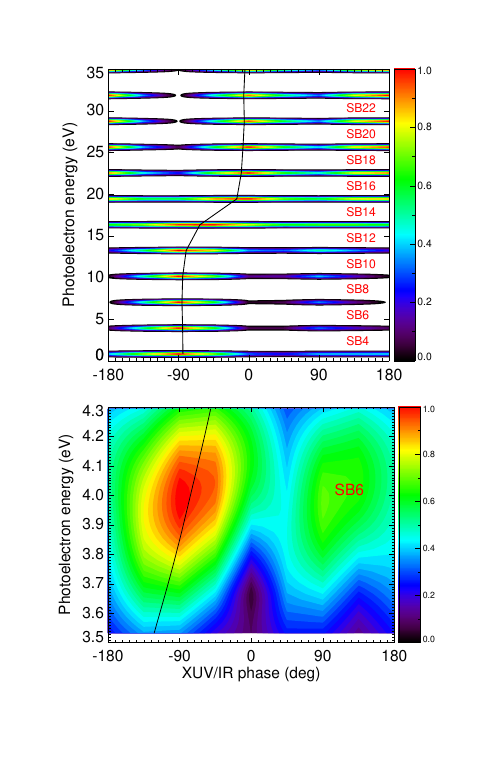}
\vs{-15mm}


\caption{ 
RABBITT trace of Li $2s$ at $\omega=1.55$~eV.
\emph{Top}: Multiple sidebands (SBs) are shown, with their orders
indicated on the right-hand vertical axis. The solid line connects the
sideband centers and serves as a guide of the ESB.
\emph{Bottom}: Expanded view of SB6. The solid line indicates the
ISB. The horizontal axis on both panels shows the relative XUV--IR phase.  }
\label{Fig2}
\end{figure}

\begin{figure}[t]

\includegraphics[width=1.0\columnwidth]{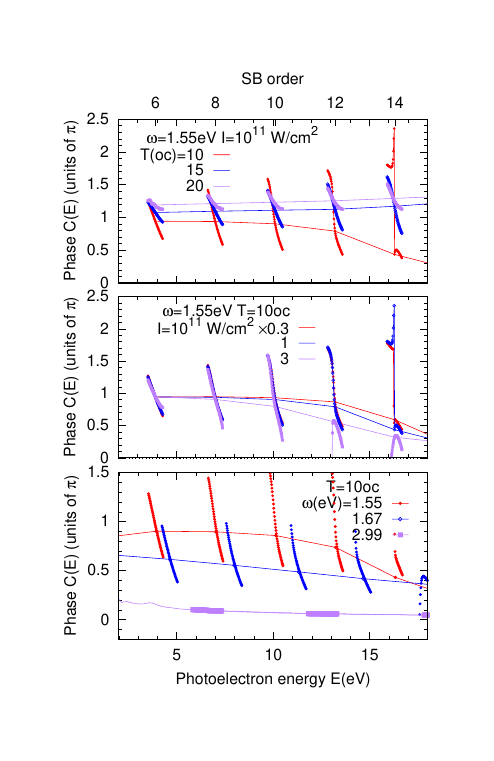}
\vs{-15mm}



\caption{ RABBITT phase parameter $C(E)$ as a function of the IR
  pulse duration $T_{\rm IR}$ (top), the IR peak intensity $I$ (middle) and IR frequency
  $\w$ (bottom).  In all panels the ISB and ESB phases are shown with
  similarly colored dots and solid lines, respectively.
}
\label{Fig3}
\end{figure}

\paragraph*{Results.}

Figure~\ref{Fig2} shows the RABBITT trace for lithium in
the $2s$ state at $\omega=1.55$~eV ($\Delta\approx-0.12$~eV in the model
potential).
The upper panel is the conventional peak-resolved RABBITT trace; the
ESB phase $C$ (solid blue curve) sweeps nearly $\pi$ between SB4 and SB22,
consistent with Ref.~\cite{PhysRevA.104.L021103}.
That behavior is background context here.

The lower panel is the defining new observable: the rainbow-resolved
phase $C(E)$ inside a single sideband (SB6).  The phase of the
delay-dependent oscillation --- traced by the diagonal black line
marking the cosine maximum --- varies by nearly $\pi$ across the
$\sim0.8$~eV width of the sideband (see also Fig.~S1 of the
Supplemental Material).  No single time delay can describe this
energy-resolved phase; it is a \emph{phase landscape} imprinted by the
Rabi dynamics.

The physical origin is the interference, at energy $E$, between
(i) the two-photon amplitude from the unperturbed $2s$ state and
(ii) the one-photon amplitude from the IR-populated $2p$ state,
weighted by $a_p(t_k)$ at the arrival time of each APT pulselet.
Because $a_p(t_k)$ changes across the APT (the $2p$ population builds
up during the IR leading edge), the effective ionization envelope of the
$2p$ channel is non-Gaussian and time-varying, imprinting a non-trivial
phase profile $C(E)$ in the energy domain via \Eref{eq:Mr}.

The ESB phase sign reversal with detuning, reported
in Ref.~\cite{PhysRevA.104.L021103} and reproduced here (see
Supplemental Material), together with the close
agreement between rRABBITT and crRABBITT far from
resonance, confirms that the ISB structure in
Fig.~2 is a genuine pulsed-APT effect rather than a
numerical artifact.

\Fref{Fig3} presents the central quantitative analysis of the ISB
phase $C(E)$ over a range of SB orders. The figure displays the $C(E)$
dependence on the IR pulse duration $T_{\rm IR}$ (top), the IR peak intensity
$I$ (middle) and the IR frequency $\omega$ (bottom).

The top panel displays $C(E)$ vs.\ photoelectron energy within each
SB at $\omega=1.55$~eV for three IR pulse durations of
10, 15, 20 optical cycles (oc).
Two trends are immediately evident.
First, the ISB phase profile \emph{sharpens} with increasing SB order:
the phase sweep concentrates into a narrower energy window near the
sideband center.
This is explained by \Eref{eq:Mr}: the resonant-to-non-resonant
amplitude ratio $r_{2q}\propto[\sigma_{2p}(E_{2q})/\sigma_{2s}(E_{2q})]^{1/2}$
decreases with SB order because $\sigma_{2p}\propto E^{-9/2}$ falls off
faster than $\sigma_{2s}\propto E^{-7/2}$~\cite{PhysRevA.104.L021103}.
As $r_{2q}$ decreases, the ISB phase contrast is confined to the narrow
peak of the sideband where the resonant channel is still comparably
strong, producing the observed sharpening.

Second, the ISB phase \emph{flattens} as $T_{\mathrm{IR}}$ increases.
This behavior reflects the long-pulse limit (i): when the IR pulse
duration greatly exceeds the APT, all pulselets sample the same
instantaneous $a_p$, the time-varying phase structure averages away,
and $\delta C(E)\to\phi_r$ becomes energy-independent.  Conversely,
for short $T_{\mathrm{IR}}$ the Rabi cycle is unfinished across the
APT duration, successive pulselets sample $a_p$ at different Rabi
phases, and the resulting coherent sum has a rich energy-dependent
phase via \Eref{eq:Mr}.

The middle panel displays $C(E)$ within each SB at $\omega=1.55$~eV
and $T_{\rm IR}=10$~oc for the three IR peak intensities $I(10^{11}{\rm
  W/cm^2})=0.3$, 1, and 3.  Increasing the intensity enhances the
resonant contribution and increases the overall phase excursion, but
leaves the spectral slope of the ISB profile nearly unchanged.  This
behavior indicates that the spectral shape of the ISB dispersion is
governed primarily by the temporal phase structure of the Rabi-dressed
wave packet, whereas the overall phase excursion remains sensitive to
the strength of the resonant coupling.  In contrast, varying the pulse
duration in the top panel directly modifies the temporal interval over
which successive APT pulselets sample the Rabi evolution, leading to a
substantial change of the ISB slope.

The bottom panel shows $C(E)$ at fixed $T_{\mathrm{IR}}=10$~oc while
varying $\omega_{\mathrm{IR}}$ from 1.55~eV toward, through and away
from the resonance.  As $\omega\to\omega_{sp}$, the ISB phase profile
flattens.  This behavior reflects the common-mode cancellation
discussed above: at $\Delta=0$ the dressed amplitudes $a_s$ and $a_p$
maintain a fixed $-\pi/2$ relative phase for any pulse area, so the
resonant dressing acts as a common gate on both RABBITT arms without
shifting their relative phase.  The ISB phase modulation is maximized
near $|\Delta|\approx\Omega$, where $\varphi_s(t)$ of \Eref{eq:argAs}
is largest, and vanishes both at $\Delta=0$ and for
$|\Delta|\gg\Omega$.  RABBITT thus measures the \emph{dressed-state
  dynamical phase}, not the population inversion, and these two
quantities peak at \emph{different} values of $\Delta$ --- a
counterintuitive but experimentally verifiable distinction.

\paragraph*{Discussion.}

\emph{ISB phase as a Rabi-cycle clock.}  Equations~(\ref{eq:Mrf}) and
(\ref{eq:dC}) show that the ISB phase is essentially the argument of
the Fourier transform of $a_p(t)\cdot f_{\mathrm{APT}}(t)$ evaluated
at the energy offset $E-E_{2q}$ from the sideband center.  The
Rabi-cycle parameters $(\Omega_D, T_{\mathrm{IR}}, \Delta)$ control
the time-domain shape of $a_p(t)$, and the spectral phase of this
product directly maps onto $C(E)$.  Three observables together
constitute a self-consistent set of signatures of the Rabi parameters:
(a) the width of the ISB phase feature is sensitive to $\Omega_D$;
(b) the \emph{amplitude} of the
phase sweep encodes $\Delta/\Omega_D$ (vanishes at resonance, peaks at
$|\Delta|\approx\Omega$); (c) the \emph{SB-order dependence} of
sharpening encodes the relative photoionization cross-sections
$\sigma_{2p}/\sigma_{2s}$ as a function of energy.

When a strong resonant channel is present, the standard interpretation
of the RABBITT phase as an atomic time delay via
$\tau_a=(dC/dE)/(2\omega)$~\cite{Dahlstrom2012} fails entirely.
The resonant phase enters only one arm of the RABBITT interferometer
(the $2p$ ionization path) and is not symmetric under
$\omega\to-\omega$, so the finite-difference formula cannot be applied.
The ISB phase makes this failure \emph{directly visible}: when $C(E)$
sweeps through $\pi$ within a single sideband, no self-consistent time
delay exists.

Near resonance, the harmonic peaks themselves acquire a delay-dependent
yield, with oscillation phases approximately opposite to those of the
neighboring sidebands.
This reflects coherent redistribution between the resonant $2p$ channel
and the non-resonant $2s$ channel: the one-photon harmonic amplitude
from the IR-populated $2p$ state beats against the direct $2s$ amplitude.
Standard perturbative RABBITT, which treats harmonic peaks as
delay-independent, cannot account for this effect.


The ISB phase mechanism is not specific to lithium.  Any system where
a valence-shell resonance lies near the IR probe frequency is a
candidate: other alkali atoms (Na, K, Rb, Cs) whose $ns\!\to\!np$
transitions fall in the NIR, or molecules with low-lying electronic
resonances accessible to typical Ti:sapphire driving fields.  Recent
theoretical and experimental studies of Rabi-driven photoelectron
phase structures and Autler--Townes dynamics~\cite{Mao2023,Liao2024}
support the generality of this approach.  The $2p_{m=1}$ magnetic
substate of lithium does not couple to $2s$ under linear
polarization~\cite{PhysRevA.104.L021103}, providing a non-resonant
reference within the same atom and under identical experimental
conditions for differential measurements.

Attosecond beamlines are increasingly coupled to cold-atom
setups~\cite{DeSilva2020,DeSilva2021}, making the proposed measurement
timely and feasible with existing technology.
Efforts to combine RABBITT with cold lithium in a magneto-optical trap
are underway \cite{Fischer2025}, and although technically challenging, such an
experiment would directly image the detuning-dependent ISB phase
landscape.

\paragraph*{Conclusion.}

We have identified the intra-sideband (ISB) phase dispersion in rainbow
RABBITT as a previously  unexplored  probe of coherent Rabi dynamics.
The ISB phase is governed by the differential dynamical phase of the
Rabi-dressed bound-state wave packet sampled by successive attosecond
pulses, described analytically by the Fourier transform of the
time-dependent dressed-state amplitude over the APT envelope
(Eqs.~(\ref{eq:Mrf}) and (\ref{eq:dC})).
Three experimentally accessible signatures --- sharpening with SB order,
flattening with IR pulse duration, and vanishing at exact resonance ---
form a self-consistent fingerprint of the Rabi-cycle parameters.
The counterintuitive on/off-resonance contrast (maximum ISB phase at
finite detuning, flat phase at exact resonance) distinguishes the
dynamical phase probe from population-based observables.
Rainbow RABBITT thereby acts as a Rabi-cycle clock, opening a new
window on coherent light--matter interaction in dressed quantum systems
and extending attosecond spectroscopy beyond noble-gas targets.

\begin{acknowledgments}
This work was supported by the Australian Research Council Discovery
grant DP230101253.  Computational resources of the NCI Australia were
utilized in this work. V.V.S. thanks the ANU for hospitality.
\end{acknowledgments}

\clearpage
\np

\newcommand{\ii}{\mathrm{i}}
\renewcommand{\ee}{\mathrm{e}}
\newcommand{\dd}{\mathrm{d}}

\setcounter{equation}{0}
\setcounter{figure}{0}

\renewcommand{\thefigure}{S\arabic{figure}}
\renewcommand{\thetable}{S\arabic{table}}
\renewcommand{\theequation}{S\arabic{equation}}

\section{Supplemental material}

\paragraph{Purpose of this note}

This Supplemental Material describes the numerical implementation used
for the rainbow RABBITT calculations reported in the main text.  The
main manuscript focuses on the physical interpretation of the
intra-sideband (ISB) phase in terms of the Rabi-dressed bound-state
wave packet.  Here we give the computational details: the effective
one-electron lithium model, the laser fields, the TDSE propagation,
the projection of the resonant $2s$ and $2p$ populations, and the
extraction of the energy-resolved RABBITT phase $C(E)$.

\paragraph{Single-active-electron TDSE for lithium}

The lithium atom is treated in the single-active-electron (SAE)
approximation.  The active electron evolves in an effective central
potential $V_{\rm eff}(r)$ representing the frozen ionic core.  The
field-free Hamiltonian is
\begin{equation}
  \hat H_0 = -{1\over 2}\nabla^2 + V_{\rm eff}(r) .
  \label{eq:H0_SAE}
\end{equation}
We use the optimized effective potential of 
\citet{Sarsa2004163}, which reproduces the low-lying lithium spectrum
sufficiently accurately for the present purpose.  In the model used
here the $2s\rightarrow2p$ transition energy is $\omega_{sp}=1.67~{\rm
  eV}$, compared with the experimental value $1.85~{\rm eV}$
\cite{NIST_ASD}.  The detuning of the IR field is therefore defined as
$ \Delta=\omega-\omega_{sp}$ where $\omega$ is the IR carrier
frequency.

The time-dependent Schr\"odinger equation is solved in the combined
XUV and IR fields,
\begin{equation}
  \ii {\partial\over\partial t}\Psi({\bf r},t)
  = \left[\hat H_0 + \hat W(t)\right]\Psi({\bf r},t),
  \label{eq:TDSE_supp}
\end{equation}
with the dipole interaction written in the length gauge,
\begin{equation}
  \hat W(t)=-z\,E(t),
  \qquad
  E(t)=E_{\rm XUV}(t-\tau)+E_{\rm IR}(t).
  \label{eq:field_total}
\end{equation}
Both fields are linearly polarized along the $z$ axis and $\tau$ is
the XUV--IR delay.  The propagation was carried out with the TDSE
solver developed by \citet{PhysRevA.84.062701} within the Paraxial
Approximation and time-dependent Hartree--Fock (PAHF) framework.  The
same numerical implementation has been benchmarked and successfully
applied in earlier studies of attosecond photoionization and rainbow
RABBITT processes~\cite{Serov2026circular,Serov2026rainbow}. In the
present work, the wave function was represented using a finite-element
discrete-variable representation for the radial coordinate and a
spherical-harmonic expansion for the angular dependence. Time
propagation employed an implicit fourth-order scheme, while exterior
complex scaling was used to suppress reflections from the radial box
boundary.

During the propagation, the field-free $2s$ and $2p$ populations are
monitored by projection,
\begin{equation}
  P_{2s}(t)=\left|\langle 2s|\Psi(t)\rangle\right|^2
, \ 
  P_{2p}(t)=\sum_m\left|\langle 2p_m|\Psi(t)\rangle\right|^2 .
  \label{eq:populations_supp}
\end{equation}
For linearly polarized fields and an initially prepared $2s$ state,
the dominant resonantly coupled component is $2p_{m=0}$.  These
projections provide a direct numerical check of the Rabi dynamics
discussed analytically in the main text. Convergence of the extracted
ISB phases with respect to the numerical propagation parameters was
carefully verified, and no significant changes were observed upon
further refinement of the computational grid.  

\paragraph{Laser fields}

The calculation uses an attosecond pulse train (APT) together with a
phase-locked IR pulse.  The APT is centered at $ \omega_x = 15\omega$,
and contains odd harmonics from H5 to H25 with a Gaussian spectral
envelope.  The delay-dependent XUV field can be written schematically
as
\begin{eqnarray}
  \label{eq:APT_field}
  E_{\rm XUV}(t-\tau)
  &=& f_{\rm APT}(t-\tau)\\
&&\hspace*{-2cm} \times   \sum_n E_{2n+1}
    \cos\left[(2n+1)\omega(t-\tau)+\phi_{2n+1}\right],
\nonumber
\end{eqnarray}
where $f_{\rm APT}$ is the APT envelope and $\phi_{2n+1}$ are the
harmonic phases.  The detailed harmonic phases are not essential for
the detuning dependence discussed in the main text; 
they contribute only a common group-delay term to the fitted RABBITT phase.

The IR field is taken in the form
\begin{equation}
  E_{\rm IR}(t)=E_0 f_{\rm IR}(t)\cos(\omega t),
  \label{eq:IR_field}
\end{equation}
with a $\cos^2$ temporal envelope $f_{\rm IR}(t)$ of total duration
$T_{\rm IR}$.  Calculations were performed for several IR frequencies
near the model $2s\rightarrow2p$ resonance and for several pulse
durations, typically $T_{\rm IR}=10$, $15$, and $20$ optical cycles.
This parameter range allows us to compare below-resonant, resonant,
and above-resonant driving, and to follow the crossover between the
short-pulse and long-pulse regimes described in the main text.

\paragraph{Photoelectron spectra and rainbow phase extraction}

For each value of $\omega$, $T_{\rm IR}$, and delay $\tau$, the 
angular integrated photoelectron spectrum $S(E,\tau)$ is extracted
from the propagated wave packet after the laser pulses.  In the
present lithium calculation, the relevant observable is the
energy-resolved yield in the sideband region.  The conventional
RABBITT analysis would integrate $S(E,\tau)$ over the whole sideband
and fit the resulting sideband yield to a single cosine.  In the
rainbow analysis, the same fit is performed independently at each
photoelectron energy $E$.

Eight equally spaced delays spanning one full
$2\omega$ RABBITT period were used. This choice
provides a complete discrete Fourier decomposition of
the delay-dependent signal while minimizing the total
number of TDSE propagations. Equivalently, for
$x=\omega\tau$, the signal at each energy is fitted to
\begin{equation}
  S(E,\tau)=A(E)+B(E)\cos\left[2\omega\tau+C(E)\right],
  \label{eq:fit_supp}
\end{equation}
or, in discrete Fourier form,
\begin{align}
  X(E)&=\sum_j S(E,\tau_j)\cos(2\omega\tau_j),
  \\
  Y(E)&=\sum_j S(E,\tau_j)\sin(2\omega\tau_j),
  \label{eq:Fourier_components}
\end{align}
\ms\mms
with

\begin{equation}
  C(E)=\tan^{-1}\left[-{Y(E)\over X(E)}\right], \
  B(E)\propto\sqrt{X^2(E)+Y^2(E)} .
  \label{eq:C_B_supp}
\end{equation}
This procedure isolates the $2\omega$ RABBITT component and is robust
against residual oscillations at $\omega$ or $3\omega$.  
The resulting energy-dependent phase $C(E)$ is the
intra-sideband (ISB) phase shown and discussed in the
main text. The same phase-retrieval procedure was
also applied to non-resonant calculations, where it
reproduces the conventional sideband phases obtained
from spectrally integrated RABBITT analysis, thereby
providing an independent validation of the numerical
extraction algorithm.

The sideband-integrated, or extra-sideband (ESB),
phase is obtained by first integrating the signal
over the sideband window and then applying the same
Fourier extraction procedure,
\begin{equation}
  S_{2q}(\tau)=\int_{\rm SB_{2q}} S(E,\tau)\,\dd E .
  \label{eq:ESB_supp}
\end{equation}
The comparison between the ESB phase and the intra-sideband phase
$C(E)$ demonstrates which phase structures are hidden by conventional
spectral integration.

\ms\mms\mms
\paragraph{Continuous-RABBITT benchmark}

In the continuous\\ RABBITT (crRABBITT) benchmark calculation, the
attosecond pulse train was replaced by a single isolated attosecond
pulse having the same central frequency and a comparable spectral
bandwidth.  The same TDSE propagation and Fourier phase-extraction
procedure were then applied to the resulting photoelectron spectra. In
this limit, the discrete sampling of successive APT pulselets is
absent, and the calculation therefore provides a useful reference for
distinguishing genuine pulse-train effects from artifacts of the
numerical analysis.
The agreement between rRABBITT and crRABBITT away
from the resonant condition confirms that the
pronounced intra-sideband phase structure reported
in the main text is not a numerical artifact of the
Fourier extraction procedure. Figure~S1 presents a
broader comparison of the ESB/RABBITT,
crRABBITT, and rRABBITT phases over an extended
photoelectron energy range for three representative
IR frequencies.

\begin{widetext}
\begin{figure*}[t]
\ms
\includegraphics[width=\textwidth]{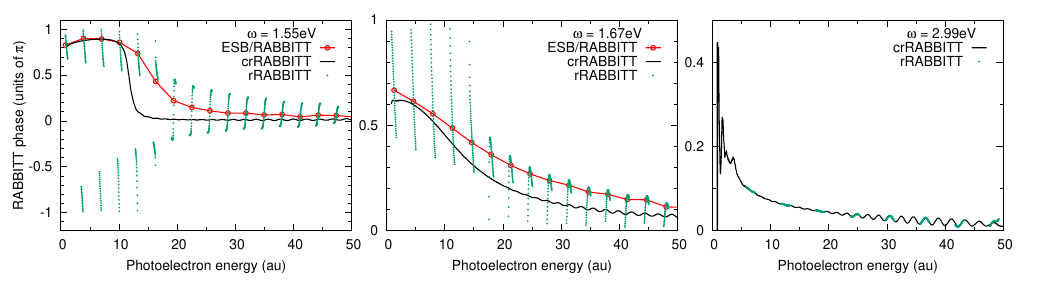}


\caption{ Comparison of the conventional sideband-integrated RABBITT
  (ESB), continuous-RABBITT (crRABBITT), and rainbow RABBITT
  (rRABBITT) phases over an extended photoelectron energy range for
  three representative IR frequencies: below resonance
  ($\omega=1.55$~eV, left), at the model $2s\to2p$ resonance
  ($\omega=1.67$~eV, center), and well above resonance
  ($\omega=2.99$~eV, right).  The conventional RABBITT and crRABBITT
  calculations yield smooth energy-dependent phase variations, whereas
  the rainbow analysis resolves the intra-sideband phase structure
  hidden by spectral integration.  The pronounced oscillatory
  modulation observed below resonance is progressively suppressed as
  the IR frequency approaches and then moves away from the resonance.
  The weak residual structure visible in the right panel originates
  from under-threshold Rydberg excitations, analogous to those
  observed in the uRABBITT regime~\cite{Serov2026circular}.  } \ms
\end{figure*}
\end{widetext}

\vs{-10mm}
\paragraph{Relation to the analytical model}

The numerical phase extraction described above can be compared directly
with the analytical expression used in the main text.  The resonant
amplitude contains a coherent sum over ionization from the
IR-populated $2p$ state,
\begin{equation}
  \widetilde M_r(E;\Delta,T_{\rm IR})
  \propto
  \int_{-\infty}^{\infty}
  \dd t\, f_{\rm APT}(t)\,a_p(t)\,
  \ee^{\ii(E-E_{2q})t},
  \label{eq:Mr_supp}
\end{equation}
where $a_p(t)$ is the excited-state amplitude generated by the IR
field.  Thus the rainbow phase $C(E)$ is sensitive to the spectral
phase of the time-dependent resonant amplitude.  
%


\end{document}